\input{aipcheck}

\documentclass[
    ,final            % use final for the camera ready runs
  ]
  {aipproc}

\layoutstyle{6x9}

\begin{document}

\title[Further Observations of the Intermediate Mass Black Hole Candidate ESO 243-49 HLX-1]
      {Further Observations of the Intermediate Mass Black Hole Candidate ESO 243-49 HLX-1}
      


\author{S. A. Farrell}{
  address={Department of Physics \& Astronomy, University of Leicester, University Road, Leicester, LE1 7RH, UK},
  email={saf28@star.le.ac.uk},
  %thanks={This work was commissioned by the AIP}
}
\author{M. Servillat}{
  address={Harvard-Smithsonian Center for Astrophysics, 60 Garden Street, MS-67, Cambridge, MA 02138, USA},
}
\author{S. R. Oates}{
  address={Mullard Space Science Laboratory/UCL, Holmbury St. Mary, Dorking, Surrey, RH5 6NT, UK},
}
\author{I. Heywood}{
  address={University of Oxford, Department of Physics, Keble Road, Oxford, OX1 3RH, UK},
}
\author{O. Godet}{
  address={Universit\'e de Toulouse, UPS, CESR, 9 Avenue du Colonel Roche, F-31028 Toulouse Cedex 9, France},
}
\author{N. A. Webb}{
  address={Universit\'e de Toulouse, UPS, CESR, 9 Avenue du Colonel Roche, F-31028 Toulouse Cedex 9, France},
  %email={arno@mittelbach-online.de},
}
\author{D. Barret}{
  address={Universit\'e de Toulouse, UPS, CESR, 9 Avenue du Colonel Roche, F-31028 Toulouse Cedex 9, France},
}

\begin{abstract}
The brightest Ultra-Luminous X-ray source HLX-1 in the galaxy ESO 243-49 currently provides strong evidence for the existence of intermediate mass black holes. Here we present the latest multi-wavelength results on this intriguing source in X-ray, UV and radio bands. We have refined the X-ray position to sub-arcsecond accuracy. We also report the detection of UV emission that could indicate ongoing star formation in the region around HLX-1. The lack of detectable radio emission at the X-ray position strengthens the argument against a background AGN.
\end{abstract}

\maketitle

%%%%%%%%%%%%%%%%%%%%%%%%%%%%%%%%%%%%%%%%%%%%
%% MAINMATTER
%%%%%%%%%%%%%%%%%%%%%%%%%%%%%%%%%%%%%%%%%%%%

\section{Introduction}

The brightest ultra-luminous X-ray source (ULX) currently known (HLX-1) was discovered in the second {\it XMM-Newton} Serendipitous Source Catalogue \citep[2XMM;][]{wat09} in the
outskirts of the edge-on spiral galaxy ESO 243--49 \citep{far09}. Its 0.2 -- 10 keV unabsorbed X-ray luminosity, assuming the galaxy distance (95 Mpc), exceeded $1.1 \times 10^{42}$ ergs s$^{-1}$. Follow-up observations have revealed large scale luminosity and spectral variability in X-rays, in a way similar to Galactic black hole binary systems \citep{god09}. The extreme luminosities of ULXs, if the emission is isotropic and below the Eddington limit, imply the presence of an accreting black hole with a mass of $\sim$10$^2$ -- 10$^5$ M$_\odot$. However, the existence of such intermediate mass black holes (IMBHs) is in dispute. Both super-Eddington accretion and beaming of the X-ray emission could account for X-ray luminosities up to $\sim10^{40}$ergs s$^{-1}$ for stellar mass black holes (10 -- 50 M$_\odot$), but would require extreme tuning to explain an X-ray luminosity of $10^{42}$ ergs s$^{-1}$. Hence, HLX-1 is an excellent candidate IMBH \citep{far09}. The existence of IMBHs has profound implications for massive star evolution, and the formation and evolution of star clusters and galaxies in general \citep[e.g.][]{mil04}. In this paper we present the results of new and archival multi-wavelength observations of HLX-1 in an attempt to confirm an association with ESO 243-49 and determine in particular where it lies within the galaxy (e.g. in a star cluster, star forming region, globular cluster etc.).

\section{X-ray Data}

We obtained a 1 ks observation of HLX-1 under the Director's Discretionary Time (DDT) program with the HRC-I camera onboard {\it Chandra} on 2009 July 4 (ObsID: 10919). No source was detected within the {\it XMM-Newton} error circle of HLX-1, indicating the count rate must have dropped down to $<$ 0.006 cts s$^{-1}$, compared to the expected 0.03 cts s$^{-1}$ based on the most recent \emph{XMM-Newton} flux and spectrum measurements \citep{far09}. \emph{Swift} XRT monitoring observations of HLX-1 confirmed the drop in flux and found that one month later the flux increased significantly \citep{god09}. Following this re-brightening we obtained a second deeper DDT observation of 10 ks with the HRC-I  on 2009 August 17 (ObsID: 11803). A total of 11 sources were detected, including a source consistent with the position of HLX-1 with a net count rate of 0.098 $\pm$ 0.003  cts s$^{-1}$. After applying  astrometry correction by cross-matching detected sources against the 2MASS catalogue \citep[see][for a full discussion]{web09}, a final position of RA = 01h 10m 28.29s, Dec = -46$^\circ$ 04' 22.3" was obtained for HLX-1, with a 95\% error of 0.3".

\section{UV Data}

\begin{figure}
\caption{\emph{Swift} UVOT \emph{uvw2} image of ESO 243-49. The white contours show the orientation of the galaxy. The black circle indicated by the white arrow is centered on the {\it Chandra} position of HLX-1, with the radius representing the 95$\%$ error bounds}
\includegraphics[width=0.55\textwidth]{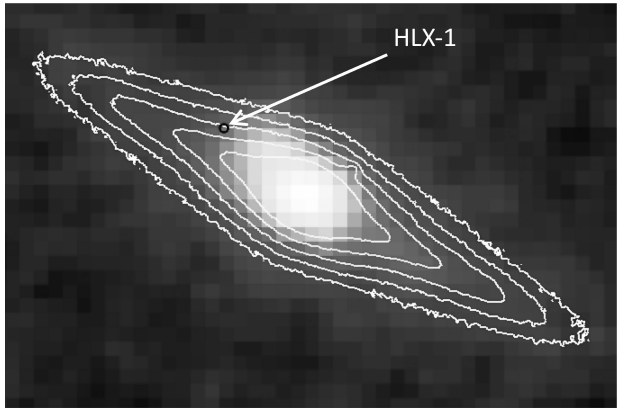}\label{uvot}
\end{figure}

The {\it Swift} UVOT observed the field of ESO 243-49 on 2009 August 5, 6, 16,
18, 19 and 20 for a total exposure of 38~ks.
Observations were performed in the {\it uvw2} ($\sim$160 -- 250 nm) filter only. At the location
of the core of ESO 243-49 there is an extended object (Figure \ref{uvot}), with some hints of an elongated emission towards the position of HLX-1. No point source is observed above the flux level of the
galaxy at the {\it Chandra} position of HLX-1, although given the spatial resolution of 2.9" FWHM in the {\it uvw2} band\footnote{http://heasarc.nasa.gov/docs/heasarc/caldb/swift/docs/uvot/} we cannot rule out a point source unresolved from the nuclear emission. We estimate a 3$\sigma$
upper-limit of $20.3$ mag at this position \citep[see][for details]{web09}. 

\begin{figure}
\caption{Archival \emph{GALEX} NUV (left) and FUV (right) images of ESO 243-49. The white contours show the orientation of
the galaxy. The black circles indicated by the white arrows are
the \emph{Chandra} position of HLX-1, with the radii indicating the 95\% error bounds.}
\includegraphics[width=\textwidth]{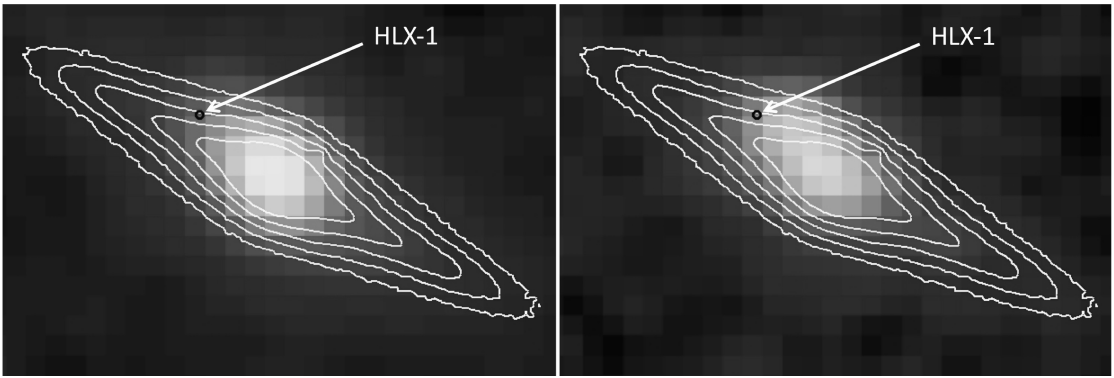}\label{galex}
\end{figure}

The field of HLX-1 was observed by
{\it GALEX} as part of the deep survey on 2004 September 27 for $\sim$13 ks in the
near-UV (NUV, $\sim$180 -- 280 nm) and $\sim$8 ks in the far-UV (FUV, $\sim$150 -- 200 nm). A clear extension towards the location of
HLX-1 from the nucleus of ESO 243-49 can be seen in both images (Figure
\ref{galex}), with the dominant emission occurring in the FUV. No point source was
detected coincident with the position of HLX-1 in either band, although again an unresolved point source cannot be ruled out due to the spatial resolution \citep[4 -- 6" FWHM;][]{mor05}. 3$\sigma$ upper limits of
20.4 mag and 21.4 mag were determined in the NUV and FUV respectively \citep[see][for details]{web09}. 

\section{Radio Data}

The field of ESO 243-49 was observed as part of the Phoenix Deep Survey (PDS) with the Australia Telescope Compact Array at 1.4 GHz \citep{hop03}. The final mosaic and a series of source catalogues are
in the public domain\footnote{http://www.physics.usyd.edu.au/$\sim$ahopkins/phoenix} and we have used these data products to search for radio emission from HLX-1. A source is clearly detected coincident with the nucleus of ESO 243-49, but no source is present at the position of HLX-1. The flux density at the location of HLX-1 is $\sim$5~$\mu$Jy. This is below the 3$\sigma$ local rms noise level of 45~$\mu$Jy, which we thus adopt as the flux density upper limit for HLX-1. We fitted a Gaussian to the radio source at the nucleus of ESO 243-49, yielding an integrated flux density of 0.15 $\pm$ 0.03~mJy. The derived position is consistent with the PDS position of RA = 01h~10m~27.69s, Dec = -46$^{\circ}$~04'~27.8'' with uncertainties of 0.2'' and 0.5'' respectively. Given the accurate astrometry in both radio and X-ray it is extremely unlikely that the 0.15~ mJy source detected in the nucleus of ESO 243-49 is associated with HLX-1. The spatial resolution of the PDS observations is high enough such that any significant radio emission at the position of HLX-1 would be distinct from that arising from the nucleus of ESO 243-49.

\section{Conclusions}

Using \emph{Chandra} observations of HLX-1, we have
determined an improved position with a 95\% confidence
error radius of 0.3". Coincident with this new position there appears
to be some evidence for possibly extended UV emission. Although
we cannot definitively say that this emission
is related to HLX-1, the fact that no similar
extended emission is seen in the radio, infrared,
optical or X-ray domains \citep[][Webb et al. in prep.]{far09}, makes it unlikely that
this is either a foreground or background source. The fact that
this emission appears to be stronger in the FUV
could hint towards star formation taking place in
that region, as the UV flux primarily originates
from the photospheres of O- through later-type
B-stars (M $>$ 3 M$_\odot$), and thus measures star formation averaged over a 10$^8$ yr timescale \citep[e.g.][]{lee09}. Alternatively,
if the UV emission is truly extended it could indicate the presence of a trail of star formation created
by ram-pressure stripping of material from a
dwarf galaxy that has recently interacted with
ESO 243-49 \citep[e.g.][]{sun09}. In this case HLX-1 could be
an IMBH which was once
at the centre of the dwarf galaxy, but has now had
most of the gas and stars stripped from it via the
gravitational interaction with ESO 243-49. The lack of radio emission down to 45~$\mu$Jy rules out a blazar \citep{tur07} and strengthens the argument that a background AGN is not the source of the X-ray emission.

%%%%%%%%%%%%%%%%%%%%%%%%%%%%%%%%%%%%%%%%%%%%%%%%
%% BACKMATTER
%%%%%%%%%%%%%%%%%%%%%%%%%%%%%%%%%%%%%%%%%%%%%%%%

\begin{theacknowledgments}
S.A.F. and S.R.O. acknowledge STFC
 funding. O.G. acknowledges funding from the CNRS and CNES. M.S. is supported in part by \emph{Chandra} grant AR9-0013X. We thank Harvey Tananbaum and the \emph{Chandra} team as well as Neil Gehrels and the \emph{Swift} team for according us and supporting the DDT and ToO observations. This research has made use of data obtained from the
{\it Chandra} Data Archive and software provided by the {\it Chandra} X-ray Center. We acknowledge the use of public data from the \emph{Swift} data archive. We thank the {\it GALEX} collaboration and the Space Telescope Science Institute
for providing access to the UV images used in this work. {\it GALEX}
is a NASA Small Explorer Class mission. The Australia Telescope Compact Array is part of the Australia Telescope which is funded by the Commonwealth of Australia for operation as a National Facility managed by CSIRO.
\end{theacknowledgments}



\bibliographystyle{aipproc}   % if natbib is available

\bibliography{farrell_final}


\end{document}